\documentclass[]{article}
\usepackage{natbib}

\usepackage{footnote}
\usepackage{times}
\usepackage{xspace}
\usepackage{url}

\usepackage{amssymb}
\usepackage{amsmath}
\usepackage{amsthm}
\usepackage{enumitem}
\usepackage{graphicx}
\usepackage{relsize}
\usepackage{layouts}
\usepackage{verbatim}
\usepackage{multirow}
\usepackage{colortbl}
\usepackage{tcolorbox}

\usepackage[colorinlistoftodos]{todonotes}

\newcommand{\con}[1]
{\textsf{#1}\xspace}
\newcommand{\bfo}{BFO\xspace}

\usepackage{authblk}

\title{Measuring Expert Performance at Manually Classifying \\Domain
  Entities under Upper Ontology Classes}

\author[1]{Robert Stevens}
\author[2]{Phillip Lord}
\author[3]{James Malone}
\author[1]{Nicolas Matentzoglu}

\affil[1]{School of Computer Science, University of Manchester, Oxford Road, M13 9PL, Manchester, United Kingdom}
\affil[2]{FactBio, Innovation Centre, Cambridge Science Park, CB4 0EY, Cambridge, United Kingdom}
\affil[3]{School of Computing, Newcastle University, 1 Science Square,
NE4 5PG, Newcastle-upon-Tyne, United Kingdom}

\affil[]{robert.stevens@manchester.ac.uk,
  phillip.lord@newcastle.ac.uk,
  james@factbio.com, nicolas.matentzoglu@gmail.com}

\begin{document}

\maketitle




%
%
%

\begin{abstract}

\emph{Background.}
Classifying entities in domain ontologies under upper ontology classes is a recommended task in ontology engineering to facilitate semantic interoperability and modelling consistency. Integrating upper ontologies this way is difficult and, despite emerging automated methods, remains a largely manual task.

\emph{Problem.}
Little is known about how well experts perform at upper ontology
integration. To develop methodological and tool support, we first need to understand how well experts do this task. We designed a study to measure the performance of human experts at manually classifying classes in a general knowledge domain ontology with entities in the Basic Formal Ontology (BFO), an upper ontology used widely in the biomedical domain.

\emph{Method.}
We recruited 8 BFO experts and asked them to classify 46 commonly known entities from the domain of travel with BFO entities. The tasks were delivered as part of a web survey.

\emph{Results.}
We find that, even for a well understood general knowledge domain such as travel, the results of the manual classification tasks are highly inconsistent: the mean agreement of the participants with the classification decisions of an expert panel was only 51\%, and the inter-rater agreement using Fleiss' Kappa was merely moderate (0.52). We further follow up on the conjecture that the degree of classification consistency is correlated with the frequency the respective BFO classes are used in practice and find that this is only true to a moderate degree (0.52, Pearson).

\emph{Conclusions.}
We conclude that manually classifying domain entities under upper ontology classes is indeed very difficult to do correctly.
Given the importance of the task and the high degree of inconsistent classifications we encountered, we further conclude that it is necessary to improve the methodological framework surrounding the manual integration of domain and upper ontologies.
\end{abstract}

\section{Introduction}
Upper ontologies, sometimes called upper-level ontologies, conceptualise basic categories of entities common across all domains, such as physical entities and processes, and can serve as tools facilitating semantic interoperability and data integration~\cite{DBLP:journals/sigmod/Noy04}. Upper ontology integration is the task of selecting appropriate super or equivalent classes in an upper ontology for a given entity in a domain ontology. Integrating domain ontologies with upper ontologies is a difficult task so, despite the emergence of automated methods~\cite{schmidt2016analysing}, is often performed manually. To date, little is known of how, and how well, experts perform at upper ontology integration.

We investigate how well experts in the Basic Formal Ontology (BFO)~\cite{Fricke17}, an upper ontology used widely in the biomedical domain, can manually classify domain entities under BFO classes. This is, to our knowledge, the first systematic study to determine the performance of experts at upper ontology integration. Beyond a detailed profile of expert performance, our work forms the first stage in the development of a technique that should make it easier to perform upper ontology integration consistently correctly within the \bfo, and, by extension, any other upper ontology.

\emph{Upper ontologies} seek to distinguish all the categories, common across domains, under which entities can be classified~\cite{DBLP:journals/sigmod/Noy04}. Apart from the BFO, the focus of our work, a number of ontologies have been suggested as upper ontologies~\cite{mascardi_comparison_2006}, including the Descriptive Ontology for Linguistic and Cognitive Engineering (DOLCE)~\cite{DBLP:conf/ekaw/GangemiGMOS02}, the General Formal Ontology (GFO)~\cite{Herre2010} and the Suggested Upper Merged Ontology (SUMO)~\cite{DBLP:conf/fois/NilesP01}.

The advocates of upper ontologies argue that they afford a consistent, correct modelling style~\cite{DBLP:journals/ao/SmithC10}--therefore as a consequence, different ontologies developed according to the same upper ontology can be integrated. Others have used upper ontologies to improve the performance of ontology matching algorithms~\cite{mascardi_automatic_2010}. The Open Biomedical Ontologies (OBO) foundry~\cite{PMID:17989687} recommends the use of BFO as an upper ontology. Given that all the OBO ontologies are represented from the same ontological perspective, then the many OBO ontologies should be able to be drawn together without impediment. Thus all OBO ontologies that have entities represented as `role', for example, are roles from a \bfo perspective and are integrated safely as subclasses of `role'. What relationships hold between entities is also determined by the upper ontology (for example, a process can be part of another process, but not part of a continuant, and a continuant participates in a process), so conforming to the same upper ontology means relationships are used consistently. Much of this point of view is assertion, but there is some evidence that the use of upper ontologies improves the consistency of modelling~\cite{DBLP:conf/esws/Keet11}.

Whether or not the utility of upper ontologies is true is not relevant to this study; that they are used in ontology development is relevant~\cite{DBLP:conf/coopis/BorgoL04,DBLP:conf/esws/Keet11}. There is anecdotal evidence that upper ontologies are difficult for many ontology developers to use. This represents a significant barrier both to contributing to an ontology and in understanding how an ontology has been structured. We aim to make \bfo, and other upper ontologies, easier to use.

To integrate a domain ontology with an upper ontology, an ontologist not only needs to have a good knowledge of the domain she is attempting to represent but also of the classes in the upper ontology. There are many articles describing upper ontologies~\cite{DBLP:conf/woa/MascardiCR07}, and some see upper levels and the approach they represent as an ontology method~\cite{DBLP:journals/ao/SmithC10,DBLP:series/ihis/GuarinoW09}. Despite that, integrating upper ontologies remains a challenging and error-prone task. One important reason for that is the difficulty of understanding the, often highly abstract, definitions of entities in the upper ontology. For example, the BFO defines a continuant as `an entity that persists, endures, or continues to exist through time while maintaining its identity', and an entity \emph{b} as a `\emph{specifically dependent continuant} where, b is a continuant and there is some independent continuant c which is not a spatial region and which is such that b s-depends on c at every time t during the course of b’s existence'. While these definitions may work for some well-trained specialists, it should be obvious that the average domain expert may not be able to derive an intuition from these definitions that is strong enough to correctly classify a particular domain entity underneath it.

We see a need for an accessible and usable method or instrument for domain specialists that typically build ontologies to be able to use upper ontologies correctly. If we are to develop an instrument to aid people using an upper ontology, that technique should be at least as effective as experienced users of that upper ontology. We also need to understand where difficulties in using an upper ontology lie. Through our survey, we show that classifying general knowledge domain entities is hard, even for experts of the BFO, and establish a performance baseline for future method or instrument development. To determine the potential impact of a learning effect, we correlate expert performance with the degree to which entities are used across a well-known ontology repository.

We present our analysis of the following research questions:

\begin{enumerate}
\item How well do \bfo experts classify domain entities under BFO classes?
\item How does the particular area (occurrent, spatial region, etc.) of the \bfo affect the performance?
\item How does the self-rated confidence of the BFO experts affect the correctness of manual entity classification?
\item What distinctions in \bfo are used in practice?
\item Does \bfo usage in practice correlate with performance?
\end{enumerate}

\section{Materials and Method}
The primary goal of our study is to determine how well BFO experts classify entities in a domain ontology with entities in the \bfo. In outline, our method is as follows:

\begin{enumerate}
\item Choosing a set of general knowledge domain entities, such as `taxi', or `airplane flight', that span a significant proportion of \bfo classes.
\item Producing descriptions of those entities.
\item Creating a survey that asks people to read the descriptions, select the most appropriate super-class in the \bfo, rate the confidence with which that selection was made and give feedback in the form of comments.
\item Analysing the classifications of domain entities with upper ontology classes quantitatively regarding correctness according to a rich set of metrics and identifying particular areas of difficulty in the BFO.
\item Analysing, qualitatively, the comments regarding thoughts, concerns and difficulties as perceived by the participants.
\end{enumerate}

\subsection{Survey Design}
We downloaded BFO 1.1 from \url{http://ifomis.uni-saarland.de/bfo/} on 7 December 2014. We chose to use BFO, as it is one of the upper ontologies most widely used in the biomedical domain, and in particular \bfo version 1.1, which benefited from more wide-spread use and familiarity at the time of this study than the newer \bfo~2.0.\footnote{\url{https://github.com/BFO-ontology/BFO/wiki/BFO-1.1-to-2-changes}} Considering our long-term goal to provide methodological and tool support for the process of upper ontology integration in general, any upper ontology could have been used; familiarity and widespread use was what drove our choice to use version 1.1 of \bfo.

We selected 23 out of 39 classes in BFO, mostly leaf classes,\footnote{22 leaf classes, and 1 branch class (spatial region). We left out the following 3 leaf classes: fiat process part, scattered temporal region, temporal instant. We found it challenging to think of convincing examples for these, and this helped reduce the overall size of the questionnaire.} for inclusion into our survey. For all except two BFO classes, we selected 2 general knowledge entities from the domain of travel which we believe to be well-known to a wide range of people. We selected only 1 entity for BFO class \con{function} and 3 for \con{generically dependent continuant}\footnote{We found it challenging to find examples for \con{function} that were not suggestive of function in their description}.
The entities were drawn from diverse sub-domains such as geography, transport, persons, accommodation, activities all related to travel or holidays.

The survey was constructed using SelectSurvey.NET from classapps.\footnote{\url{https://selectsurvey.net/}} A copy of the survey used in this study is included in the supplementary materials. We collected the following demographic data: willingness to share data publicly, name, preferred email address, gender, age and self-rated level of expertise w.r.t BFO 1.1. The main survey consisted of 46 survey items (one survey item per domain entity), each comprising a description, a confidence rating scale and a comment box. Each survey item description was written in the form `[An] \emph{x} as in \ldots', where \emph{x} is the name of an entity from the travel domain, followed by a description of that entity, for example:

\begin{quotation}
\emph{drinking a beer} as in `I like drinking a beer with my pizza.'
\end{quotation}


We attempted to write each entry in sufficient detail to allow participants to make a judgment as to under which BFO entity it should be classified, but not in such a way that participants were told the answer; so, for instance, we avoided the use of words that are used in labels for entities or properties in \bfo. The draft survey was run by \bfo experts as a pilot and revisions made. We used their feedback to improve the entities chosen, the names of those entities and the description of those entities. We did not impose any restrictions on the participants regarding the use of external tooling (such as consulting the BFO in the ontology editor Prot\'{e}ge\'{e}).

Each entry in the survey was followed by:
\begin{itemize}
\item A list of the \bfo~1.1 leaf classes from which users could choose. This included an `unknown' choice.
\item A five-point Likert-scale (1-5) by which participants could report their confidence for the chosen entity.
\item A free text field for comments.
\end{itemize}

This is an example of a complete survey item:
\begin{tcolorbox}
\emph{air space} as in `We flew the airspace of three countries before we arrived.'

[Dropdown with BFO classes]

\vspace{1em}

\emph{How confident are you in the correctness of your previous answer? (1 = no confidence; 5 = totally confident)}

Confidence ()1 ()2 ()3 ()4 ()5

\vspace{1em}

Comments: [Textfield for comments]
\end{tcolorbox}

We sent invitations to the contributors to \bfo~1.1 as listed on \url{http://ifomis.uni-saarland.de/bfo/}, excluding any people that had participated in the pilot for this investigation. In the following, we will refer to the study participants as \emph{experts}, due to their role as BFO experts, and to the authors except for NM as the \emph{experimenters}, due to their role in the design and execution of the study. Ethical approval for the study was obtained from the School of Computer Science Ethics committee (University of Manchester), approval number CS 272b.

\subsection{Quantitative Analysis of Classification Correctness}
\label{sec:methquan}
The main goal of the analysis was to quantify how well participants classified domain entities under BFO classes. We define expert performance here as the degree to which the resulting classifications are correct. Given a domain entity D and a BFO entity B, we define the degree of correctness as the distance of B to the most specific superclass of D in BFO. As classes in BFO are mutually exclusive, the correct BFO entity should be uniquely identifiable. However, while there can only be a single most specific super-class, determining which one is a subjective process that can depend on the degree to which there is a common understanding of the domain entity in question, or, to a hopefully lesser extent, the shared understanding of entities in the upper ontology. For example, a layer cake (e.g., a Victoria Sponge) could be classified as either an object or an object aggregate; there is a correct decision, but it is straightforward to see why confusion may occur. Similarly, is the `soldier’ aspect of a soldier ant a function or a role?

As we have, therefore, no way to determine correctness formally, we resort to two groups of metrics that serve as proxies: \emph{Experimenter-expert agreement} is the degree to which the study participants (experts) agree with the classifications determined by the authors (experimenters) and \emph{inter-expert agreement} is the degree to which the study participants agree amongst themselves. Where possible, our measures are presented from 0 to 1, with 0 indicating no agreement (i.e. correctness), and 1 indicating total agreement (i.e. correctness). Unless stated otherwise, we use the mean for all aggregations (entity-level and overall). Table~\ref{tab:metricstable} provides an overview of the metrics we discuss in detail in the following section.

\begin{table*}[h]
  \centering
    \begin{tabular}{llp{6.5cm}p{6.6cm}}
    \multicolumn{1}{l}{Group} & Abbr. & Metric  & Purpose \\
    \\ \hline \\
       \multirow{2}[0]{*}{EE} & EEA   & Experimenter-expert classification agreement & Conformance with author classification  \\
          & EES   & Experimenter-expert classification similarity & Conformance with author classification considering ontological similarity \\
          \\ \hline  \\
    \multirow{5}[0]{*}{IE} & MAC   & Most agreed classification ratio     & Conformance with majority vote \\
          & IES   & Mean inter-expert classification similarity & Agreement considering ontological similarity \\
          & SE   & Shannon's entropy & Measure of uncertainty  \\
           & DA    & Number of different answers & Spread of the classifications \\
          & IER   & Inter-expert classification reliability & Measure for inter-rater reliability \\
          \\ \hline  \\

    \multirow{2}[0]{*}{OTH} & CON   & Mean confidence & Self-rated degree of confidence for classification \\
          & COV   & Coverage  & Proportion of ontologies an entity appears in \\
          & IMP  & Impact & Mean proportion of axioms in an ontology the entity appears in \\ \\ \hline
    \end{tabular}%
  \caption{Metrics for expert performance at manual upper ontology integration. IE are the inter-expert agreement metrics, EE the experimenter-expert agreement metrics and OTH other relevant metrics.}\label{tab:metricstable}%
\end{table*}%

\textbf{Metrics of experimenter-expert agreement} (EE) quantify the degree to which the study participants (experts) agree with the classifications determined by the authors (experimenters).

\emph{Experimenter-expert classification agreement (EEA):} $\frac{\#P_C}{\#P}$. The ratio of participants that selected the same BFO entity as the authors ($P_C$) to all participants ($P$). 1 indicates that all BFO experts shared our choice of classification, and 0 indicates that no BFO expert shared our choice of classification.

\emph{Experimenter-expert classification similarity (EES):} $\Sigma^{p}_{P}{(similarity(C_p,E))}\times\frac{1}{\#P}$, where $P$ is the set of participants, $similarity(A,B)$ is a function that computes the ontological similarity between two classes A and B, $C_p$ is the participants classification and $E$ the experimenters (authors) classification. The similarity between two classes $A$ and $B$ is the ratio of shared super-classes (including self) between $A$ and $B$ divided by the union of super-classes (of $A$ and $B$)~\cite{DBLP:journals/ki/AlsubaitPS16}. An EES of 1 represents total agreement, and an EES of 0 indicates that all 8 experts picked that had a similarity score of 0 (i.e., were completely dissimilar) from our own classification. The advantage of EES over EEA is that EES takes into account cases where the participants choice is `not far off' the experimenters choice, for example in the case where the participant picked a slightly more general class (e.g. \con{processual entity}) than the participant (e.g. \con{process}); the EEA metric does not distinguish between verdicts that are totally off (ontologically) and those that are off by just a bit.

\textbf{Metrics of inter-expert agreement} quantify the degree to which the study participants agree amongst themselves on the best upper ontology entity for the classification.

\emph{Most agreed classification ratio (MAC):} $max(\frac{x_C}{\#P})$, where $max()$ is a function that computes the maximum, $x_C$ is the number of times a class $C$ was selected and $\#P$ the total number of votes (i.e. participants). MAC is the highest proportion of votes for a domain entity. For example, the MAC of an entity $A$ given four classifications, $\{x,x,y,z\}$ is $\frac{2}{4}$. MAC quantifies the degree of agreement with the majority vote.

\emph{Inter-expert classification similarity (IES):} $\Sigma^{p}_{P}{(similarity(C_p,M))}\times\frac{1}{\#P}$, where $P$ is the set of participants, $similarity(A,B)$ is a function that computes the ontological similarity between two classes A and B, $C_p$ is the participants classification and $M$ the most agreed classification (majority vote). A value of 1 indicates total agreement between all the BFO experts (not necessarily the experimenter's own classification), and a value of 0 indicates that no BFO experts picked an answer. The IES works the same as the EES, just with a different assumption of what is the correct answer.

\emph{Shannon's entropy (SE):} $-\mathlarger{\Sigma_{xi}}{p(x_i)log_2p(x_i)}$. A measure of uncertainty. The range of the measure lies between 0 and 3, with 0 being absolute certainty and 3 being total disagreement (all 8 participants picked a different class). For readability, we present Shannon's entropy as a normalised score between 1 and 0, where 1 indicates absolute certainty (i.e. the original SE value of 0) and 0 indicates total disagreement (i.e. the original SE value of 3).

\emph{Number of different answers (DA):} The number of different answers per domain entity. For example, the DA of an entity $A$ given four verdicts, $\{x,x,y,z\}$ is 3.

Another metric of which we make use is the \emph{mean confidence rating (CON)}, the mean self-rated confidence the participants selected when determining the correct upper ontology entity. A value closer to 1 indicates very low confidence, while a value of 5 indicates the highest degree of confidence. The two remaining metrics, coverage and impact, will be discussed later in this section.

We aggregate our metrics on three different levels:
\begin{itemize}
\item Domain entity level for quantifying the correctness of classifications by domain entity, such as `tourist'. Metrics used are DA, IES, MAC, SE, EES and EEA. This allows us to talk about the difficulty of the classification of an individual domain entity.
\item Upper ontology entity level for quantifying the correctness of classifications by BFO entity, such as `role'. For lack of a formally verifiable correct classification, we assume the experimenter classification to be the correct one. Metrics are aggregated using the mean from the domain entity level. This allows us to talk about how difficult it was to classify with particular BFO categories.
\item Top level for quantifying the overall performance. Metrics used are DA, IES, MAC, SE, EES, EEA and IER, except IER aggregated using the mean from the domain entity level. This allows us to talk about the general performance of experts across a wide range of entities and respective upper ontology classes.
\end{itemize}

The main goal of our study is to determine whether classifying domain entities in their respective BFO entity manually is generally easy or generally difficult. While the small sample space of typically 2 questions per BFO entity does not permit any form of generalisation, we believe presenting some observations aggregated by entity is still interesting from an exploratory perspective.

\subsubsection{Are classification correctness or confidence correlated with entity usage?}
It is possible that a high degree of self-rated confidence or correctness was a consequence of the experience the participants had with recognising instances of particular BFO classes. For example, some BFO classes such as \con{quality} are frequently used by experienced BFO experts, while others, such as \con{scattered spatiotemporal region}, are used less often and may, therefore, be more likely to be subject to misclassification. To account for this possible bias, we performed a prevalence analysis of a snapshot of the OBO Foundry Repository~\cite{PMID:17989687}. To quantify prevalence, we use \emph{coverage} (COV), the percentage of ontologies in a given repository (the OBO Foundry in our case) that mention a particular entity ~\cite{leo2016capturing}. For example, if \con{process} appears in 2 out of 3 ontologies, we say it has a coverage of 66.7\%. We correlate coverage with a selection of performance metrics as presented above, aggregated on upper ontology entity level to determine whether classification correctness or confidence are correlated with usage. If we assume that particular groups of classifications are not avoided just because they are hard, this, if it turns out to be true, could suggest that the more frequently an upper ontology entity is used, the more classifications with that entity will be correct.

\subsection{Qualitative analysis of comments}
\label{sec:comman}
While our survey was designed to determine how well upper ontology experts are at upper ontology integration, its results can only be suggestive of the sources of difficulty. To get at least a sense of the concerns of our study participants, we gave the opportunity of leaving comments after every performed classification. We did not restrict or guide the comments in any way. We developed a coding scheme to group the comments by the following themes:
\begin{enumerate}[label=(\alph*)]
\item Comment on the text provided that describes the entity. Often about ambiguity.
\item Explanation of decision made in the choice of BFO class, including assumptions.
\item Expressing uncertainty about the decision of the chosen BFO class.
\item General comment on BFO.
\item Possible alternative answer.
\item Comment on the survey itself, rather than the wording of an entity, for example, the overall length of the survey.
\end{enumerate}

3 authors (RS, PL, JM) reviewed all comments given and classified them according to the above themes. We will use this classification to guide our narrative and identify central themes of concern and sources of difficulty to the participants.

The full report of the analysis of the data can be seen via \url{http://rpubs.com/matentzn/upperontologyclassification}. Raw data and scripts can be found on figshare~\cite{matentzoglu_stevens_malone_lord_2017}.

\section{Results and Discussion}
\label{sec:results}
Out of the 16 people we asked to participate in this study, 8
completed the survey, 4 failed to respond to the invitation after two
reminders and 4 people declined to take part in the study. The data from those people granting access to their anonymised responses may be viewed in the supplementary materials.

The ages of the participants ranged from 37 to 54 (median 43). There were four female and four male participants. The mean self-reported expertise in \bfo was 4.625 out of 5. All four male participants and one female participant rated their \bfo expertise to be 5; the remaining 3 female participants rated themselves 4 out of 5. The survey completion time varied considerably across participants: 23, 326, 347, 87, 84, 39, 90, and 33 minutes respectively. However, as we did not require participants to complete the survey in a single session, these raw numbers serve only as a rough indication of the difficulty of the survey.

Every participant was asked to classify 46 domain entities. 1 participant did not classify 7 domain entities (either by not entering a verdict or explicitly selecting the `I don't know' option), 1 participant 6 entities, 1 participant 3, 1 participant 2, and 2 participants failed to classify 1 entity. The entities with the highest number of missing answers\footnote{Overall, there were 10 entities with missing values.} were \con{situation} (4 missing), \con{why I had problems sleeping}(4), \con{air space tomorrow} (3) and \con{time and place} (3). A further 6 entities had one missing value each.\footnote{Entities with 1 missing value: \con{area where hotel will be built}, \con{distance}, \con{my hotel room}, \con{surface of pool table}, \con{tan line}, \con{warmth}} In the following, we treat missing and `I don't know' verdicts as invalid classifications, and exclude them from the analysis.

\subsection{Expert performance over time}
\label{sec:time}
\begin{figure}[h!]
\includegraphics[width=\columnwidth]{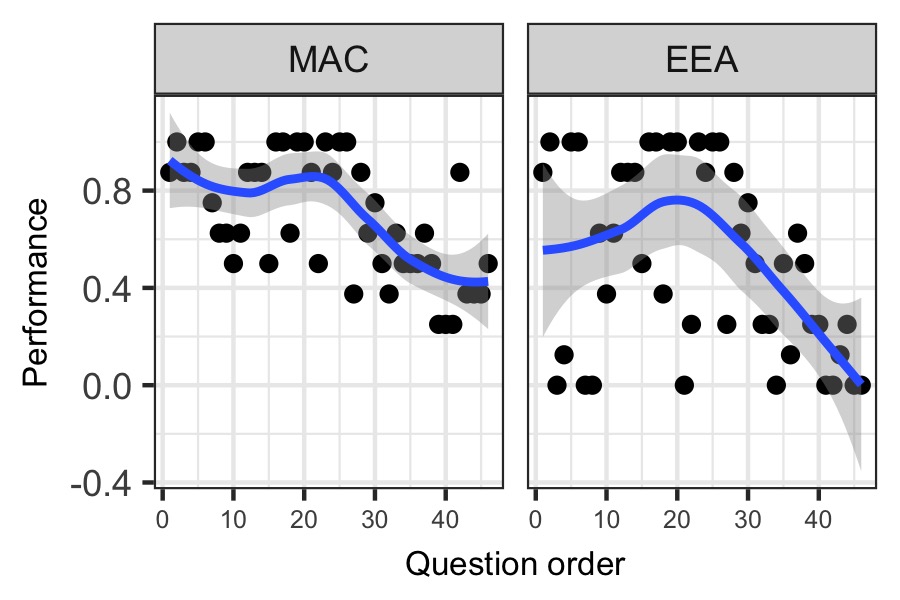}

\caption{Changes in expert classification performance over time (MAC=Most agreed classification ratio, EEA=Experimenter-expert classification agreement).}

\label{fig:conaggree}
\end{figure}

\begin{figure}[h!]
\includegraphics[width=\columnwidth]{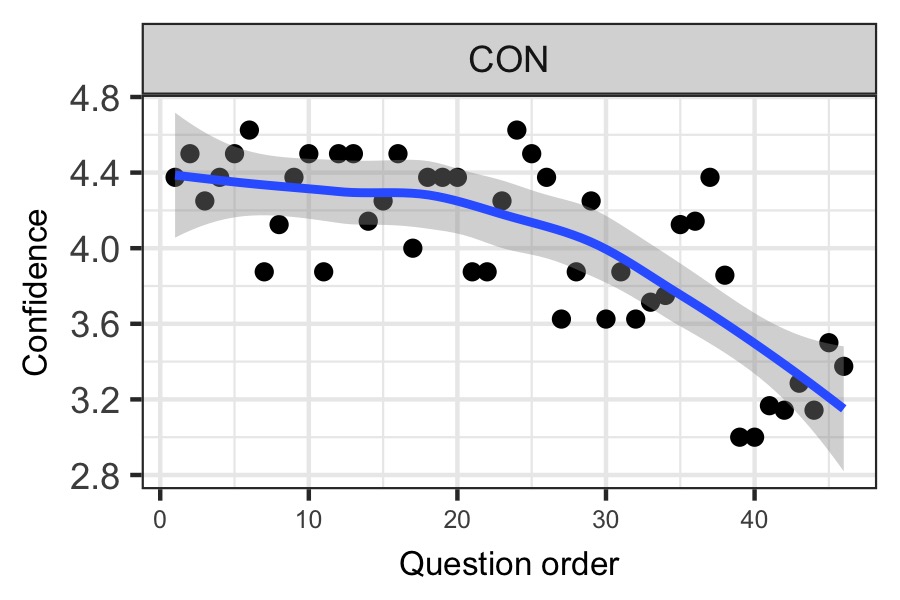}

\caption{Change in expert mean classification confidence over time.}

\label{fig:contime}
\end{figure}

We did not randomise the ordering of the questions as they were distributed to experts. Figure~\ref{fig:conaggree} shows that agreement levels go up towards the first half of the survey but then begin to fall for the last third. It is possible this is due to the survey fatigue effect, in which performance can drop due to the nature of a somewhat repetitive task. There is also the possibility, however, that the drop in performance is due to other factors, the most likely of which is that the questions became harder towards the end. This is evidenced by the participants' confidence levels falling in close correlation to the agreement scores over time (shown in Figure~\ref{fig:contime}). Additionally, we should consider that the last six domain concepts we tested all relate to `spatiotemporal' BFO-concepts (\con{spatiotemporal instant}, \con{interval} and \con{scattered spatiotemporal region}), that could simply be hard to get right. Secondly, concepts pertaining to `dimensional regions' (\con{zero}, \con{one}, \con{two} and \con{three dimensional region}) scored quite low in terms of EEA and were tested towards the last third of the survey (concept 31-38 out of 46) -- while \con{spatial region}, their direct parent concept in BFO, was tested towards the beginning of the survey (domain entities 7 and 8) and scored even lower. Lastly, out of the 9 domain entities tested in the survey where there was no agreement with the authors in terms of classification (see Section~\ref{sec:resdom}) 4 were tested in the first half of the survey, which shows that total disagreement was (almost) as likely before expert classification performance started to drop compared to after. Either hypothesis, (or indeed both) could account for some of the deterioration in agreement scores.

A possible survey fatigue could affect our results in two ways: (1) we may overestimate the difficulty of classifying domain concepts under `spatiotemporal' concepts, `dimensional regions' and \con{processual context} (the concepts tested when classification performance drops) and (2) we may underestimate the overall performance of experts at classifying domain entities under BFO concepts. Regarding (1), we report the position of each domain entity (and related BFO concept) in the survey and emphasise this potential confounder when drawing conclusions. Regarding (2): even if we had randomised the question order, aggregated results would have likely remained unchanged (randomising test order does not prevent fatigue). Furthermore, given the repetitive nature of ontology development in a realistic setting and the fact that individual ontology development sessions often exceed 90 min in our experience (6 out of 8 participants needed 90 or fewer minutes to complete the survey), fatigue will occur; therefore, preserving results that are potentially subject to survey fatigue may even add to the ecological validity (the degree to which the study and its design approximate the real-world being examined).

\subsection{BFO Expert Performance at Upper Ontology Integration}
\label{sec:resmain}
Table~\ref{tab:summarystats} shows a summary of the metrics used to quantify BFO expert performance at manual upper ontology integration. With on average (mean) 2.65 different opinions (out of 8, median: 3) regarding a classification of a domain entity and an only moderate inter-expert reliability (IER, Fleiss' Kappa) value of 0.519 (p-value 0), we can see that, even for a general knowledge domain such as travel, manual classifications can be considerably inconsistent. These results should be a warning sign, in particular when taking into account that our participants were skilled experts at the BFO---it is not inconceivable, indeed likely, that manual upper ontology integration is even more inconsistent when performed by domain experts.

As described in Section~\ref{sec:methquan}, we cannot formally determine the correctness of a classification. We, therefore, present our analysis using two different proxies for correctness. Experimenter-expert agreement metrics assume that the authors, who have had considerably more time to reflect and discuss the classification decisions than the study participants, were able to determine the correct answer. Inter-expert agreement metrics quantify how much the study participants agree among themselves. The most agreed classification ratio (MAC) in particular enables a view of the majority vote as a valid proxy for correctness. If we assume the correctness of the author's classification (EEA), we get a mere average of 51\% (median 50\%) of correct upper ontology classifications (by the study participants). If we assume the correctness of the majority vote instead (MAC), we get 69\% correct classifications (median 62\%), which is considerably better, but, given that 31\% of the classifications are wrong, still not good enough for many, if not most, cases.

The two metrics based on ontological similarity, EES and IES, draw a slightly better picture, because we consider how `far off' the participant's classifications were compared to the correct answer--in fact, they are generalisations of their respective counterparts (EEA and MAC).\footnote{EEA for example is equivalent to EES were ontological similarity between all entities is 0.} The experimenter-expert classification similarity (EES) of 0.70 looks considerably better than the corresponding exact metric EEA (0.51), which suggests that a significant number of study participant classifications were actually not `far off' the authors classification decision. The inter-expert classification similarity of 0.81 is the most lenient quantification of the correctness of the expert classifications. Even if, however, we considered 81\% correctness as acceptable in some cases, we believe the situation can, and should, be improved by methodological and tool support.

\begin{table}[!htbp]
\centering
\begin{tabular}{r|c|ccc|c}
  Metric & Mean & Min & Median & Max & SD \\
  \hline
  EEA & 0.51 & 0.00 & 0.50 & 1.00 & 0.38 \\
  EES & 0.70 & 0.14 & 0.72 & 1.00 & 0.25 \\
  \hline
  MAC & 0.69 & 0.25 & 0.62 & 1.00 & 0.25 \\
  IES & 0.81 & 0.36 & 0.82 & 1.00 & 0.17 \\
  SE & 0.67 & 0.28 & 0.65 & 1.00 & 0.24 \\
  DA & 2.65 & 1 & 3 & 5 & 1.22 \\
  \hline
  CON & 4.01 & 3.00 & 4.13 & 4.62 & 0.47 \\
\end{tabular}
\caption{Summary statistics of key metrics across all domain entities.}\label{tab:summarystats}
\end{table}

\begin{figure}[h!]
\includegraphics[width=\columnwidth]{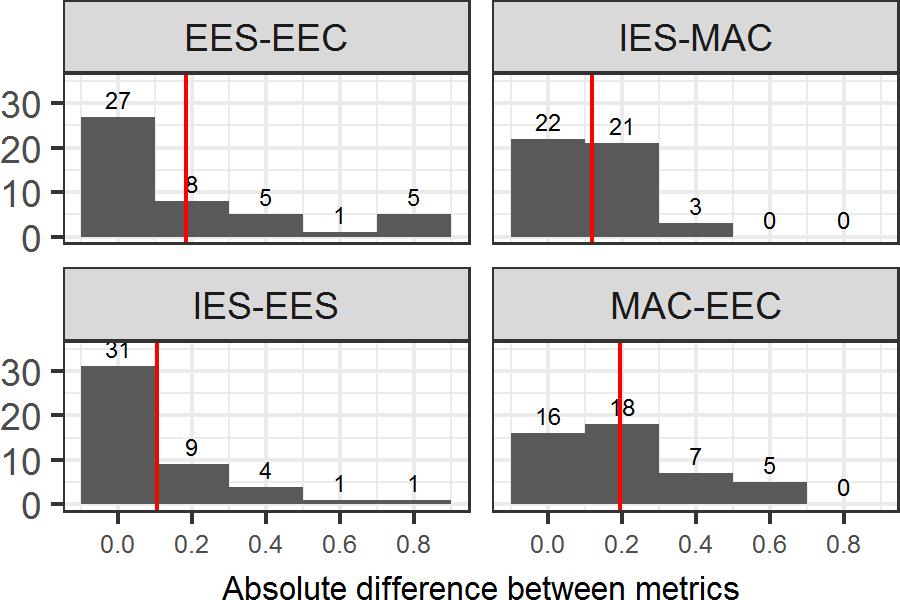}

\caption{The absolute difference between the degree of correctness across correctness metrics (y-axis: nr. of entities). EES-EEA: Difference between experimenter-expert classification similarity and agreement. IES-MAC: Difference between inter-expert classification similarity and the most agreed classification ratio. IES-EES difference between the two similarity-based metrics and MAC-EEA difference between metrics based on exact agreement. Red line indicates mean difference.}

\label{fig:EEvIE}
\end{figure}

We can see that there is a significant difference between the degree of correctness of the exact and similarity-based metrics: EEA and EES differ on average by 0.19 or 19\%, MAC and IES differ on average by 0.12 or 12\%. For a more detailed breakdown of that difference see Figure~\ref{fig:EEvIE}. While we find similarity-based metrics very informative (from an analytical perspective), the mutual exclusiveness of BFO classes suggests that to guarantee, for example, the correct functioning of applications relying on upper ontologies for data integration, we should consider exact correctness only. Figure~\ref{fig:EEvIE} also shows how much the experimenter-expert agreement (EE) metrics differ from the inter-expert agreement (IE) ones. Interestingly, there is no single case where an EE metric would suggest a more correct classification than the respective IE metric; in other words, the correctness by reference to the author's classification decision is always worse than by reference to the majority vote.

\begin{table*}[!htbp]
\centering
\begin{tabular}{r|rr|rr|rr|r}
 & EEA & EES & MAC & IES & SE & DA & CON \\
  \hline
  EEA & 1.00 & 0.91 & 0.68 & 0.63 & 0.69 & -0.65 & 0.64 \\
  EES & 0.91 & 1.00 & 0.75 & 0.75 & 0.72 & -0.68 & 0.78 \\
  \hline
  MAC & 0.68 & 0.75 & 1.00 & 0.95 & 0.96 & -0.90 & 0.69 \\
  IES & 0.63 & 0.75 & 0.95 & 1.00 & 0.87 & -0.83 & 0.71 \\
  SE & 0.69 & 0.72 & 0.96 & 0.87 & 1.00 & -0.98 & 0.65 \\
  DA & -0.65 & -0.68 & -0.90 & -0.83 & -0.98 & 1.00 & -0.65 \\
  \hline
  CON & 0.64 & 0.78 & 0.69 & 0.71 & 0.65 & -0.65 & 1.00 \\
\end{tabular}
\caption{Correlation of key metrics.}\label{tab:corrs}
\end{table*}

In the absence of a gold standard for truth, the reasons for this remain speculation. One possibility is that, in their role as BFO experts, the participants form a cluster, and are therefore more likely to agree among themselves than with ontologists outside their cluster (like the authors). Another possibility is that the authors are simply wrong in some of their classification decisions. If this was the case, it is reasonable to assume that other, non-BFO expert, ontologists would make the same, or similar, mistakes. If we make that assumption, we may prefer the experimenter-expert agreement metrics for estimating performance. A third possibility is that some of the descriptions the authors produced for the domain entities were vague and led to different mental associations among the study participants. If we were to make this assumption, we would prefer to use the inter-expert metrics to estimate classification performance, as participants at the very least all read the same text, and do not have a preconception of the meaning of the description. For this work, we leave it up to the reader to make one assumption or another; most of the analysis points to the same conclusion, namely that upper ontology integration is hard, and classification of domain entities under upper ontology classes can be highly inconsistent.

The consequences of faulty upper ontology classifications are situation-dependent. As long as a representational mistake is made `consistently' in a single ontology (e.g. \con{site} is consistently used in place of \con{spatial region}), then the consequences, for example for query answering, are probably minimal. However, ontology development efforts especially in the biomedical domain are often shared by multiple authors, and therefore more likely to suffer from inconsistent modelling. The (conceptually) same kinds of entities may appear in different branches of the ontology class hierarchy, and consequently, ontology queries may not be answered correctly. Moreover, ontology matching approaches~\cite{mascardi_automatic_2010} and data and knowledge integration applications such as the Monarch Initiative~\cite{mungall_monarch_2017} rely on integrating large numbers of biomedical ontologies (all of which are developed by different groups of experts), using upper ontologies and semantic similarity algorithms. Here, inconsistent classification is not only highly likely, but can lead to incomplete, or, often worse, wrong, answers when querying the ontologies, which can have severe consequences for example for disease diagnostics. Lastly, misclassification may lead to (large portions of) ontologies becoming unusable due to logical errors. This commonly happens when an entity is classified by two different ontologies under two mutually exclusive BFO classes (such as continuant and occurrent).

\begin{figure}[h!]
\includegraphics[width=\columnwidth]{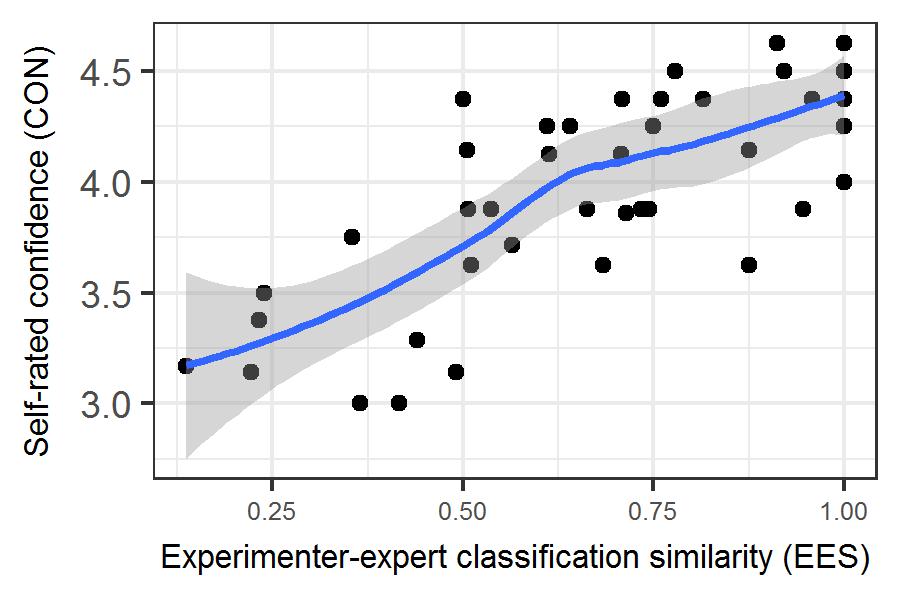}

\caption{Relationship of mean self-rated confidence (CON) and experimenter-expert classification similarity (EES).}

\label{fig:cor_con_ees}
\end{figure}

\begin{figure}[h!]
\includegraphics[width=\columnwidth]{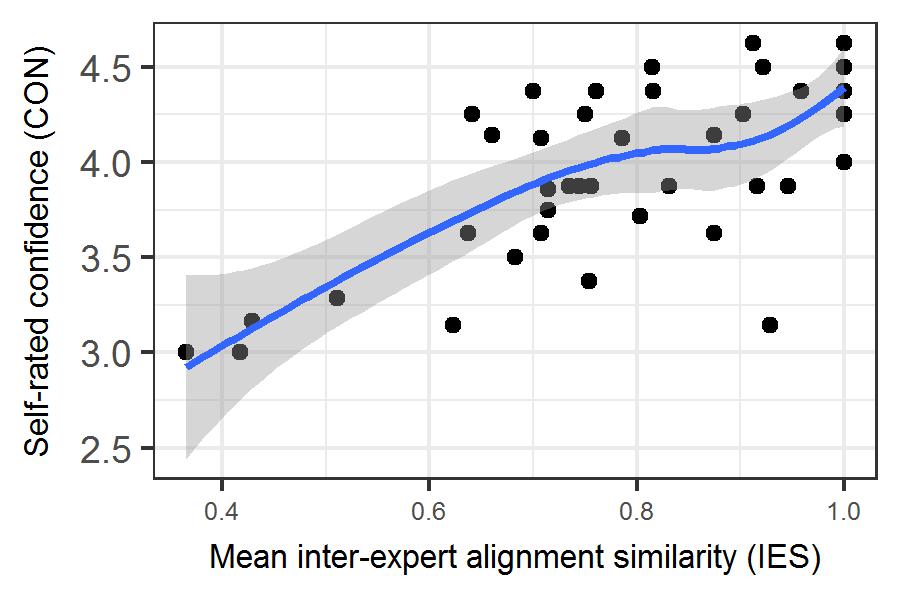}

\caption{Relationship of mean self-rated confidence (CON) and inter-expert classification similarity (IES).}

\label{fig:cor_con_ies}
\end{figure}

Table~\ref{tab:corrs} shows the correlation of all the key metrics. The first striking observation is that \emph{all} metrics are correlated.\footnote{All correlations are statistically significant at p=0.05} The weakest correlation is, with 0.64 (which we still consider strong), between self-rated confidence (CON) and experimenter-expert classification agreement (EEA). Even if we can observe some interesting differences on the level of the individual classes, we would observe similar, if not the same, tendencies on a population level. Two striking correlations are between Shannon's entropy (SE), a metric used to quantify uncertainty, and the, much simpler, metric of number of different answers (DA), of 0.98 (similarly of SE and MAC with 0.96). This suggests that SE might be redundant as a metric in scenarios like ours. Figure~\ref{fig:cor_con_ees} and \ref{fig:cor_con_ies} show how confidence correlates with the two similarity-based metrics in more detail. They suggest that those answers of which the participants lacked confidence were those that were more likely to be incorrect; this is what would be expected. These results provide further evidence for the belief that upper ontology integration is not only difficult, but that ontologists are aware that it is, and when in particular. This could help when developing tools and methods of which ontology engineers can make use whenever they feel uncertain about the correct classification of a domain entity under an upper ontology class.\footnote{Consider the opposite, where ontologists have a sense of confidence that is unrelated to the difficulty of the classification problem; they might simply not seek methodological or tool support when necessary.}

\subsection{Domain Entity-level analysis}
\label{sec:resdom}
Table~\ref{tab:mainres} shows the experimenter-expert agreement metrics for all 46 questions asked. Overall, 10 entities were classified unanimously (including the experimenter verdict), 8 entities led to two differing opinions, 12 entities to 3 opinions, 11 to 4 and 5 entities were classified by 5 different verdicts. We believe this wide spread of possible answers to be a further, strong, indication that integration with upper ontologies is error prone. For the first 10 entities, from bus driver to tourist, all participants shared the experimenter verdict. 9 entities have an experimenter-expert classification agreement ratio (EEA) of 0, which means that there was no single participant that shared the experimenter verdict. Interestingly, 7 out of these 9 entities are related to two branches of the BFO: \con{spatiotemporal region} (which is an \con{occurrent}) and \con{spatial region} (which is a continuant). Note that while we tested 4 out of these 9 concepts early on in the survey, the other five were tested towards the end, and are therefore potentially subject to the survey fatigue effect (see Section~\ref{sec:time}). Overall, the majority of the study participants disagreed with the author's classification for 19 out of the 46 domain entities, a further strong indication that consistent modelling across groups with different backgrounds (i.e. the authors and the BFO experts) is difficult.

\begin{figure*}[h!]
\includegraphics[width=0.99\textwidth]{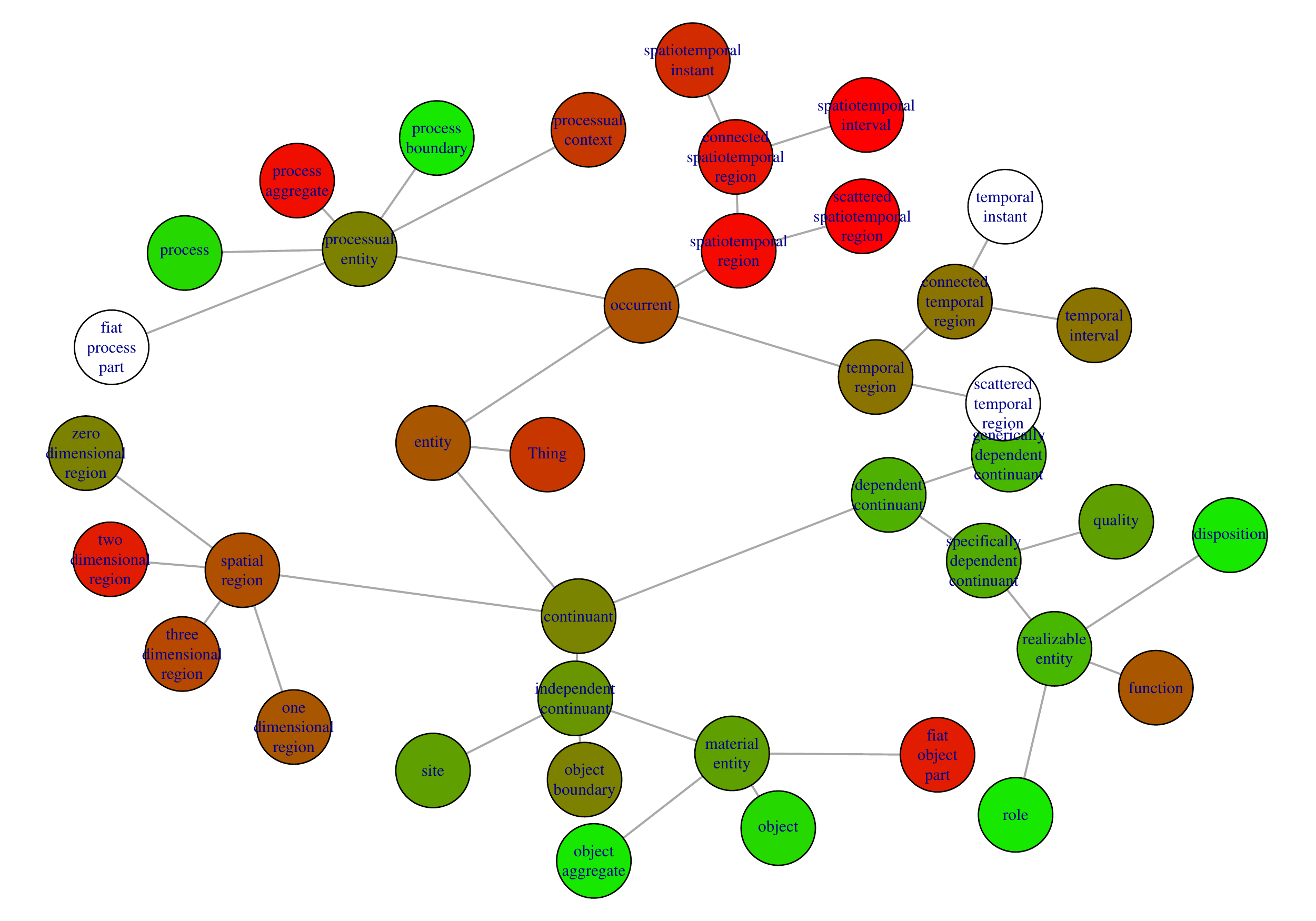}

\caption{BFO: Difficult branches analysis. Class colour indicates experimenter-expert classification agreement (EEA), aggregated as the mean of the class itself and all its children, from red (low agreement) to green (high agreement).}

\label{fig:graph_bfo_EEA}
\end{figure*}

Figure~\ref{fig:graph_bfo_EEA} shows an analysis of the difficulty of the different subparts of BFO. Using the experimenter-expert classification agreement (EEA) aggregated by the BFO class as chosen by the authors, we further aggregate each class by its EEA value \emph{and the value of its children}. For example, if \con{spatiotemporal instant} has an EEA of 1 and \con{spatiotemporal interval} one of 0.8, then \con{connected spatiotemporal region} (their parent), which does not have an EEA value of its own\footnote{I.e. participants were not asked to classify entities in that category} would have an aggregated EEA value of $\frac{0.8+1.0}{2}=0.9$. We can see that the BFO classes mentioned above, \con{spatiotemporal region} and \con{spatial region} indeed belong to the worst branches of BFO regarding classification correctness--and therefore, perhaps, to the most difficult ones to understand and get right. This is further confirmed by some of the comments we received: `Geographic locators tend to be ambiguous in language' and `there’s plenty of room for confusion with BFO spatial regions'. BFO here perhaps mirrors the difficulties shown by models from physics where both 3D and 4D models are commonly used. 

In 6 out of these 9 cases related to \con{spatiotemporal region} and \con{spatial region}, the BFO experts selected \con{site} as the appropriate alternative. We suspect that this is because the relationship between the classes of BFO here is both complex and do not reflect the etymology of the terminology. Spatial Region is, for example, a continuant, while temporal and spatiotemporal regions are occurrents, and there are no subclass relationships between any of these three. So, the spatial region, for example, `occupied by my desk' is not a subclass of the `spatiotemporal region which the desk occupies'; nor is it specified in BFO if these two are partonomic. We posit that, fundamentally, it is difficult for users to make these distinctions consistently.

\begin{table*}[!htbp]
\centering
\small
\begin{tabular}{ll|ll|ll}

Entity & SID & EES & EEA & Experimenter  & Majority vote \\
\hline
bus driver &  20 & 1.00 & 1.00 & role & role \\
  deckchairs on the beach &  25 & 1.00 & 1.00 & object aggregate & object aggregate \\
  drinking a beer &   2 & 1.00 & 1.00 & process & process \\
  end of cooking &   5 & 1.00 & 1.00 & process boundary & process boundary \\
  end of flying &   6 & 1.00 & 1.00 & process boundary & process boundary \\
  person &  23 & 1.00 & 1.00 & object & object \\
  tendency to be mosquito bitten &  17 & 1.00 & 1.00 & disposition & disposition \\
  tendency to defiate &  16 & 1.00 & 1.00 & disposition & disposition \\
  tour party &  26 & 1.00 & 1.00 & object aggregate & object aggregate \\
  tourist &  19 & 1.00 & 1.00 & role & role \\
  \hline
  airplane flight &   1 & 0.96 & 0.88 & process & process \\
  edge &  28 & 0.95 & 0.88 & object boundary & object boundary \\
  digital photograph &  13 & 0.92 & 0.88 & gen. dep. continuant & gen. dep. continuant \\
  vacation brochure &  12 & 0.92 & 0.88 & gen. dep. continuant & gen. dep. continuant \\
  bus &  24 & 0.91 & 0.88 & object & object \\
  bay &  30 & 0.88 & 0.75 & site & site \\
  warmth &  14 & 0.88 & 0.88 & quality & quality \\
  point &  37 & 0.82 & 0.62 & zero dim. region & zero dim. region \\
  take off time &  10 & 0.78 & 0.38 & temporal interval & temporal instant \\
  vacation weekend &   9 & 0.76 & 0.62 & temporal interval & temporal interval \\
  my hotel room &  29 & 0.75 & 0.62 & site & site \\
  line of lattitude &  31 & 0.74 & 0.50 & one dim. region & one dim. region \\
  timetable &  11 & 0.73 & 0.62 & gen. dep. continuant & gen. dep. continuant \\
  kilometre zero &  38 & 0.71 & 0.50 & zero dim. region & zero dim. region \\
  dining &   4 & 0.71 & 0.12 & process aggregate & \cellcolor{blue!25}process \\
  air space &  35 & 0.71 & 0.50 & three dim. region & three dim. region \\
  border &  27 & 0.68 & 0.25 & object boundary & \cellcolor{blue!25}fiat object part \\
  hotel beach &  22 & 0.66 & 0.25 & fiat object part & \cellcolor{blue!25}site \\
  distance &  15 & 0.64 & 0.50 & quality & quality \\
  \rowcolor[gray]{0.9} space &   8 & 0.61 & 0.00 & spatial region & \cellcolor{blue!25}site \\
  \rowcolor[gray]{0.9} clubbing &   3 & 0.61 & 0.00 & process aggregate & \cellcolor{blue!25}process \\
  place on wall where the postcard is put &  33 & 0.56 & 0.25 & two dim. region & \cellcolor{blue!25}site \\
  \rowcolor[gray]{0.9} tourist area &  21 & 0.54 & 0.00 & fiat object part & \cellcolor{blue!25}site \\
  tan line &  32 & 0.51 & 0.25 & one dim. region & \cellcolor{blue!25}object boundary \\
  \rowcolor[gray]{0.9} vacation location &   7 & 0.51 & 0.00 & spatial region & \cellcolor{blue!25}site \\
  area where the hotel will be built &  36 & 0.51 & 0.12 & three dim. region & \cellcolor{blue!25}site \\
  to wash &  18 & 0.50 & 0.38 & function & \cellcolor{blue!25}process \\
  moment &  44 & 0.49 & 0.25 & spatiotemporal instant & \cellcolor{blue!25}temporal instant \\
  time and place &  43 & 0.44 & 0.12 & spatiotemporal instant & \cellcolor{blue!25}spatiot. interval* \\
  situation &  40 & 0.42 & 0.25 & processual context & \cellcolor{blue!25}processual context* \\
  why I had problems sleeping &  39 & 0.36 & 0.25 & processual context & processual context* \\
  \rowcolor[gray]{0.9} surface of pool table &  34 & 0.35 & 0.00 & two dimensional region & \cellcolor{blue!25}object boundary \\
  \rowcolor[gray]{0.9} uncovered parts of the beach &  45 & 0.24 & 0.00 & scattered spatiot. reg. & \cellcolor{blue!25}fiat object part \\
  \rowcolor[gray]{0.9} patches on the floor &  46 & 0.23 & 0.00 & scattered spatiot. reg. & \cellcolor{blue!25}object aggregate \\
  \rowcolor[gray]{0.9} flood plain &  42 & 0.22 & 0.00 & spatiot. interval & \cellcolor{blue!25}site \\
  \rowcolor[gray]{0.9} air space tomorrow &  41 & 0.14 & 0.00 & spatiot. interval & \cellcolor{blue!25}site* \\

\end{tabular}

\caption{Correctness of BFO experts for classifying travel domain entities as BFO 1.1. classes. SID is survey ID or position the entity was tested as part of the survey (e.g. airplane flight was the first entity tested). EES is the experimenter-expert classification similarity. EEA is the experimenter-expert classification agreement score. The majority vote is the BFO class that was selected by the majority of BFO experts. Majority votes marked with an asterisk (*) were voted less often then there were `Not given' answers. Greyed-out lines are those with no agreement between the authors and any of the participants. Blue cells denote majority votes that differ from the authors' classifications.}\label{tab:mainres}
\end{table*}

Table~\ref{tab:mainres2} shows the breakdown of the inter-expert agreement metrics by entity. The advantage of considering inter-expert agreement metrics is that they are more objective in the sense that they are not biased by the author's preconceptions, i.e. as metrics of correctness, they do not depend on the opinions of authors.
For seven entities, the majority vote metric (MAC) deviated considerably ($>=$0.5) from the experimenter-expert agreement (EEA): \con{dining}, \con{flood plain}, \con{tourist area}, \con{clubbing}, \con{vacation location}, \con{space} and \con{surface of the pool table}. Except for \con{dining}, these belong to the same group of 9 entities identified earlier with zero agreement between the authors and the participants. For the entity clubbing, for example, the description was:

\begin{quotation}
We went clubbing every night, because I like many people dancing together.
\end{quotation}

The experimenter verdict was that \con{clubbing} was a \con{process aggregate}, rather than a \con{process}, which was the \bfo class chosen by 7 of the 8 participants (1 opted for \con{object aggregate}). One participant that classified \con{clubbing} underneath \con{process} commented on the issue suggestion `Maybe a process aggregate?', suggesting that there is, indeed, an ambiguity that requires a clear methodological protocol to overcome. The experimenters thought this was an aggregate because, while clubbing, the people dancing are involved in a set of individual processes, and not interacting with each other; however, we suspect experts felt that the clubbers are in one place, dancing to one piece of music, and thus involved in one process. It would have been possible to make this difference more explicit, but as experimenters, we faced the difficulty of not simply answering the question through the entity description.

There are three entities with majorities of size two (\con{air space tomorrow}, \con{situation} and \con{why I had problems sleeping}), all of which also have a considerable number of missing values (see the beginning of Section~\ref{sec:results}) and very low average self-rated confidence (CON) scores. \con{tan line} and \con{moment} have the most number of different opinions (5); the considerably better scores based on the similarity-based agreement metric (IES) compared to the exact metric (MAC), however, show that many of the classifications were not that `far off'. Still, we consider it interesting that such a crucial entity as `moment', no matter how ambiguous the description,\footnote{`The moment I stepped out of the door of the plane, the air smelt good'} causes such a large number of different classifications. As before, low scores may potentially be due to survey fatigue (see Section~\ref{sec:time}), as they were among the entities tested last.

\begin{table*}[!htbp]
\centering
\small
\begin{tabular}{ll|lll|ll|l}
Entity & QID & MAC & DIF & IES & SE & DA & CON \\
  \hline
  bus driver &  20 & 1.00 & 0.00 & 1.00 & 1.00 &   1 & 4.38 \\
  deckchairs on the beach &  25 & 1.00 & 0.00 & 1.00 & 1.00 &   1 & 4.50 \\
  drinking a beer &   2 & 1.00 & 0.00 & 1.00 & 1.00 &   1 & 4.50 \\
  end of cooking &   5 & 1.00 & 0.00 & 1.00 & 1.00 &   1 & 4.50 \\
  end of flying &   6 & 1.00 & 0.00 & 1.00 & 1.00 &   1 & 4.62 \\
  person &  23 & 1.00 & 0.00 & 1.00 & 1.00 &   1 & 4.25 \\
  tendency to be mosquito bitten &  17 & 1.00 & 0.00 & 1.00 & 1.00 &   1 & 4.00 \\
  tendency to defiate &  16 & 1.00 & 0.00 & 1.00 & 1.00 &   1 & 4.50 \\
  tour party &  26 & 1.00 & 0.00 & 1.00 & 1.00 &   1 & 4.38 \\
  tourist &  19 & 1.00 & 0.00 & 1.00 & 1.00 &   1 & 4.38 \\
  \hline
  airplane flight &   1 & 0.88 & 0.00 & 0.96 & 0.82 &   2 & 4.38 \\
  \rowcolor[gray]{0.9}  dining &   4 & 0.88 & \cellcolor{blue!25}0.75 & 0.96 & 0.82 &   2 & 4.38 \\
  edge &  28 & 0.88 & 0.00 & 0.95 & 0.82 &   2 & 3.88 \\
 \rowcolor[gray]{0.9}   flood plain &  42 & 0.88 & \cellcolor{blue!25}0.88 & 0.93 & 0.82 &   2 & 3.14 \\
  digital photograph &  13 & 0.88 & 0.00 & 0.92 & 0.82 &   2 & 4.50 \\
  vacation brochure &  12 & 0.88 & 0.00 & 0.92 & 0.82 &   2 & 4.50 \\
  \rowcolor[gray]{0.9}  tourist area &  21 & 0.88 & \cellcolor{blue!25}0.88 & 0.92 & 0.82 &   2 & 3.88 \\
  bus &  24 & 0.88 & 0.00 & 0.91 & 0.82 &   2 & 4.62 \\
  \rowcolor[gray]{0.9}  clubbing &   3 & 0.88 & \cellcolor{blue!25}0.88 & 0.90 & 0.82 &   2 & 4.25 \\
  bay &  30 & 0.75 & 0.00 & 0.88 & 0.65 &   3 & 3.62 \\
  warmth &  14 & 0.88 & 0.00 & 0.88 & 0.82 &   2 & 4.14 \\
  \rowcolor[gray]{0.9}  vacation location &   7 & 0.75 & \cellcolor{blue!25}0.75 & 0.83 & 0.65 &   3 & 3.88 \\
  point &  37 & 0.62 & 0.00 & 0.82 & 0.57 &   3 & 4.38 \\
  take off time &  10 & 0.50 & 0.12 & 0.81 & 0.53 &   3 & 4.50 \\
  place on wall where the postcard is put &  33 & 0.62 & 0.38 & 0.80 & 0.57 &   3 & 3.71 \\
  \rowcolor[gray]{0.9}  space &   8 & 0.62 & \cellcolor{blue!25}0.62 & 0.79 & 0.68 &   2 & 4.12 \\
  vacation weekend &   9 & 0.62 & 0.00 & 0.76 & 0.48 &   4 & 4.38 \\
  hotel beach &  22 & 0.50 & 0.25 & 0.76 & 0.42 &   4 & 3.88 \\
  patches on the floor &  46 & 0.50 & 0.50 & 0.75 & 0.42 &   4 & 3.38 \\
  my hotel room &  29 & 0.62 & 0.00 & 0.75 & 0.48 &   4 & 4.25 \\
  line of lattitude &  31 & 0.50 & 0.00 & 0.74 & 0.42 &   4 & 3.88 \\
  timetable &  11 & 0.62 & 0.00 & 0.73 & 0.57 &   3 & 3.88 \\
  \rowcolor[gray]{0.9}  surface of pool table &  34 & 0.50 & \cellcolor{blue!25}0.50 & 0.71 & 0.42 &   4 & 3.75 \\
  kilometre zero &  38 & 0.50 & 0.00 & 0.71 & 0.53 &   3 & 3.86 \\
  border &  27 & 0.38 & 0.12 & 0.71 & 0.36 &   4 & 3.62 \\
  air space &  35 & 0.50 & 0.00 & 0.71 & 0.53 &   3 & 4.12 \\
  to wash &  18 & 0.62 & 0.25 & 0.70 & 0.68 &   2 & 4.38 \\
  uncovered parts of the beach &  45 & 0.38 & 0.38 & 0.68 & 0.48 &   3 & 3.50 \\
  area where the hotel will be built &  36 & 0.50 & 0.38 & 0.66 & 0.42 &   4 & 4.14 \\
  distance &  15 & 0.50 & 0.00 & 0.64 & 0.53 &   3 & 4.25 \\
  tan line &  32 & 0.38 & 0.12 & 0.64 & 0.28 &   5 & 3.62 \\
  moment &  44 & 0.38 & 0.12 & 0.62 & 0.28 &   5 & 3.14 \\
  time and place &  43 & 0.38 & 0.25 & 0.51 & 0.40 &   4 & 3.29 \\
  air space tomorrow &  41 & 0.25 & 0.25 & 0.43 & 0.36 &   4 & 3.17 \\
  situation &  40 & 0.25 & 0.00 & 0.42 & 0.42 &   4 & 3.00 \\
  why I had problems sleeping &  39 & 0.25 & 0.00 & 0.36 & 0.42 &   4 & 3.00 \\
\end{tabular}

\caption{Inter-rater agreement of BFO experts for classifying travel domain entities. Sorted by mean inter-expert classification similarity (IES). SID: survey ID or position the entity was tested as part of the survey (e.g. airplane flight was the first entity tested). DIF: Difference of MAC to EEA, the corresponding experimenter-expert metrics. Marked cells with  DIF $>=$0.5. }\label{tab:mainres2}
\end{table*}

\subsection{Does classification correctness correlate with wide-spread use of a class?}

Table~\ref{tab:bfoclasses} shows the BFO expert classification performance and class coverage (COV), aggregated by experimenter verdict. There are only moderate correlations between class coverage and the various correctness metrics: 0.42 (SE), 0.52 (EEA), 0.49 (EES), 0.37 (IES), -0.45 (DA), 0.44 (MAC) and 0.47 (CON). The strongest correlation is between coverage and the experimenter-expert classification agreement (EEA). We found this quite surprising, as we expected there to be a strong correlation between how often a BFO class is used and the ability to use it consistently, but this is not what we found in practice. Ultimately, however, the entities that people need to use in their ontology reflect their use case, rather than their understanding of these entities--a decision, for instance, that all entities of a particular type are placed beneath a particular BFO class is made once and will, in all likelihood, be made by people expert in BFO based on use cases.

\begin{table*}[!htbp]
\centering
\begin{tabular}{rc|cc|cc|c|c}
Class & CID & EEA & EES & MAC & IES & CON & COV \\
  \hline
  quality &   7 & 0.69 & 0.76 & 0.69 & 0.76 & 4.20 & 27.94 \\
  disposition &   8 & 1.00 & 1.00 & 1.00 & 1.00 & 4.25 & 26.47 \\
  role &  10 & 1.00 & 1.00 & 1.00 & 1.00 & 4.38 & 23.53 \\
  function &   5 & 0.38 & 0.50 & 0.62 & 0.70 & 4.38 & 19.85 \\
  generically dependent continuant &   9 & 0.79 & 0.86 & 0.79 & 0.86 & 4.29 & 18.38 \\
  site &  15 & 0.69 & 0.81 & 0.69 & 0.81 & 3.94 & 14.71 \\
  spatial region &   4 & 0.00 & 0.56 & 0.69 & 0.81 & 4.00 & 13.97 \\
  object &  12 & 0.94 & 0.96 & 0.94 & 0.96 & 4.44 & 13.24 \\
  object aggregate &  13 & 1.00 & 1.00 & 1.00 & 1.00 & 4.44 & 13.24 \\
  one dimensional region &  16 & 0.38 & 0.63 & 0.44 & 0.69 & 3.75 & 12.50 \\
  three dimensional region &  18 & 0.31 & 0.61 & 0.50 & 0.68 & 4.13 & 12.50 \\
  two dimensional region &  17 & 0.12 & 0.46 & 0.56 & 0.76 & 3.73 & 12.50 \\
  zero dimensional region &  19 & 0.56 & 0.76 & 0.56 & 0.76 & 4.12 & 12.50 \\
  process boundary &   3 & 1.00 & 1.00 & 1.00 & 1.00 & 4.56 & 11.76 \\
  fiat object part &  11 & 0.12 & 0.60 & 0.69 & 0.84 & 3.88 & 10.29 \\
  object boundary &  14 & 0.56 & 0.82 & 0.62 & 0.83 & 3.75 & 10.29 \\
  process &   1 & 0.94 & 0.98 & 0.94 & 0.98 & 4.44 & 1.47 \\
  process aggregate &   2 & 0.06 & 0.66 & 0.88 & 0.93 & 4.31 & 0.74 \\
  processual context &  20 & 0.25 & 0.39 & 0.25 & 0.39 & 3.00 & 0.74 \\
  scattered spatiotemporal region &  23 & 0.00 & 0.24 & 0.44 & 0.72 & 3.44 & 0.74 \\
  spatiotemporal instant &  22 & 0.19 & 0.47 & 0.38 & 0.57 & 3.21 & 0.74 \\
  spatiotemporal interval &  21 & 0.00 & 0.18 & 0.56 & 0.68 & 3.15 & 0.74 \\
  temporal interval &   6 & 0.50 & 0.77 & 0.56 & 0.79 & 4.44 & 0.74 \\
\end{tabular}

\caption{BFO classes sorted by coverage with mean aggregated key metrics. For example, SE is the mean aggregated Shannon's entropy across all travel entities that were considered to belong to the BFO 1.1 class \con{quality} by experimenter verdict. CID represents the relative order in which a concept was tested as part of the survey (e.g.. process related entities were tested first). }\label{tab:bfoclasses}
\end{table*}

\begin{figure*}[h!]
\includegraphics[width=0.99\textwidth]{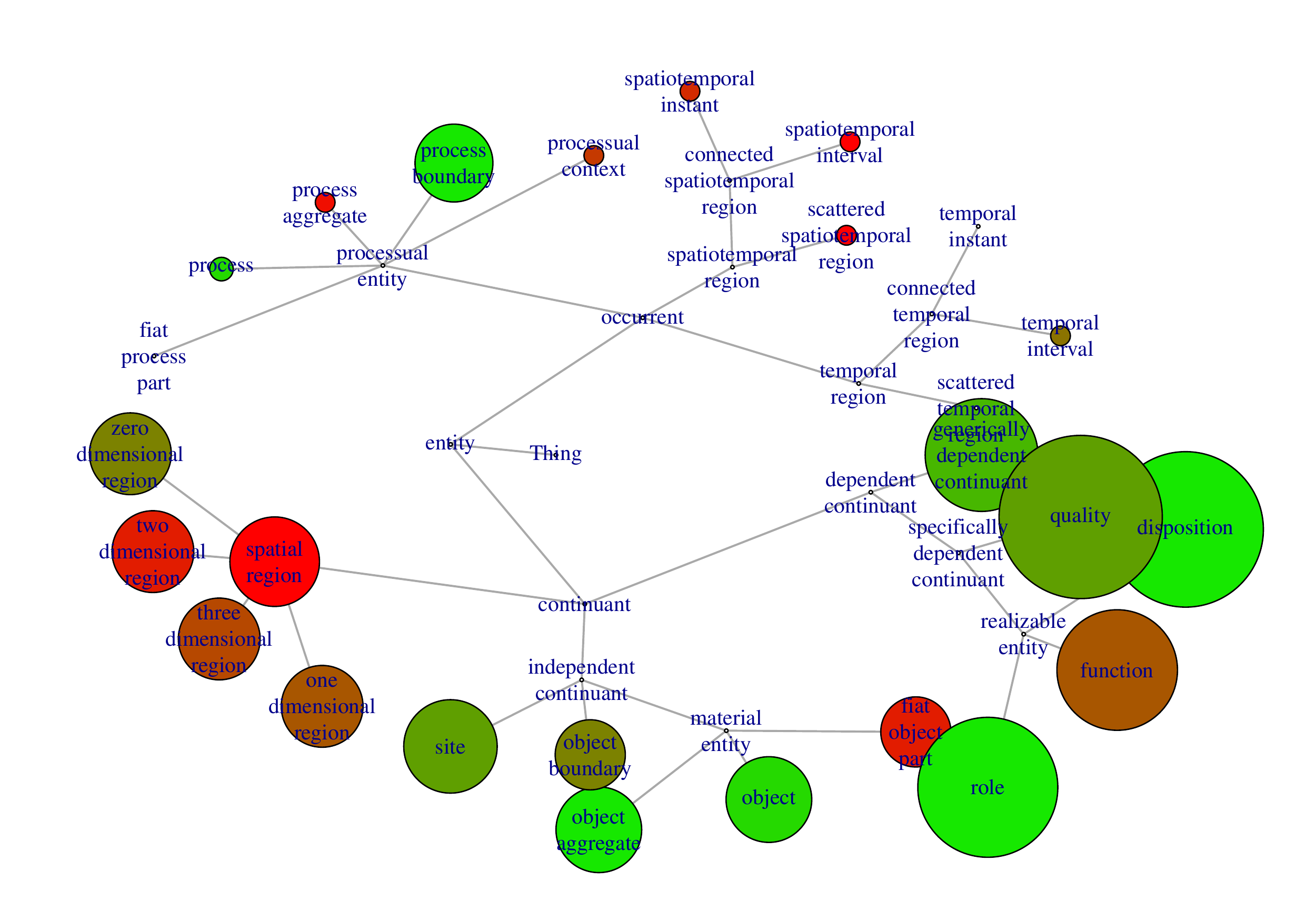}

\caption{BFO: Coverage and Difficulty. Class colour indicates experimenter-expert classification agreement (EEA), from red (low agreement) to green (high agreement). Class size indicates coverage across OBO Foundry ontologies (the higher the coverage, the larger the class node).}

\label{fig:graph_bfo_cov}
\end{figure*}

The classes with the widest coverage are \con{quality} (28\% of all parseable OBO Foundry ontologies), \con{disposition} (26\%) and \con{role} (24\%). All of these classes have indeed reasonable levels of correctness (in particular disposition and role, which have an EEA of 1), which partially confirms our initial hypothesis that widespread use and classification correctness were related, at least for popular classes. However, as we can see in Figure~\ref{fig:graph_bfo_cov}, this view cannot be confirmed in general. The fourth most used class, \con{function}, already has much lower levels of correctness (EEA 0.38), and the entire branch of \con{spatial region} looks to be decidedly more difficult than their relatively widespread use would suggest. Given the already discussed moderate correlations, we conclude that in general, widespread use is not a guarantee for more correct classifications. One possible reason is that classifying an entity in BFO is just hard.

In Section~\ref{sec:resdom}, we found that \con{site} was a popular BFO class for classifying domain entities by the study participants, preferred over the various sub-classes of \con{spatiotemporal region} and \con{spatial region} that the authors chose. One (partial) reason for this could be the relative popularity of using \con{site} (appearing in 14.7\% of the ontologies in the OBO Foundry), at least compared to the various sub-classes of \con{spatiotemporal region} (which appeared much more rarely, in less than 1\% of the ontologies). The classes around \con{spatial region} are, however, considerably more frequently used, but still less often than \con{site}. Another possibility, however, is that we, the authors, simply misclassified. One comment demonstrating the difficulty of this classification was given regarding \con{vacation location}: `It is a \con{site} rather than a \con{spatial region} {[author's choice]} because BFO’s notion of \con{spatial region} is that of a {[N]}ewtonian fixed space and on that view earth and places on earth move through different parts of space moment to moment.' Ultimately, Newtonian fixed space is not reflective of reality, and BFO does not define its frame of reference; these kinds of classifications are contentious and difficult to resolve.

\subsection{Qualitative analysis of comments}
\label{sec:commres}
In the following, we will present a qualitative analysis of the comments participants gave on each of the 46 entities in the survey. Overall, we received 109 comments - about 30\% of all items (by participant) were commented upon (109/(8*46)=0.30). The most commented items were \con{border} (5 comments), as well as \con{bay}, \con{hotel beach}, \con{take off time}, \con{tan line}, \con{tourist area}, \con{vacation location}, \con{why I had problems sleeping} (all 4 comments). We mapped the predominant themes described in Section~\ref{sec:comman} across all comments given. We determined the predominant themes through majority voting, i.e. the theme, or themes in the case of ties, that most reviewers assigned to a comment. Table~\ref{tab:comments} shows the number of comments falling into each of the categories.\footnote{Note that, because we allowed multiple themes per comment, the numbers do not add to 109.}

\begin{table}[!htbp]
\centering
\begin{tabular}{l|p{5.7cm}|l}
CID & Theme & \# \\
  \hline
A & Comment on the text provided that describes the entity. Often about ambiguity &  20 \\
B & Explanation of decision made in the choice of BFO class, including assumptions. & 52 \\
C & Expressing uncertainty about the decision of BFO class. & 12 \\
D & General comment on BFO. & 18 \\
E & Possible alternative answer. & 26 \\
F & Comment on the survey itself, rather than the wording of an entity, for example, the overall length of the survey. & 4 \\

\end{tabular}
\caption{Number of times a theme was mentioned across comments.}\label{tab:comments}
\end{table}

Some example comments from each category are:

\begin{description}
\item \emph{A--Comment on the text provided that describes the entity:} `some uncertainty here again because there are alternative interpretations. If the speaker means that there is something about themselves that gives rise to the mosquito biting, then they would be talking about a disposition. However, they could also be referring to parts of their history - the aggregate of cases when they were bitten during past trips'.
\item \emph{B--Explanation of decision made in the choice of BFO class:} `A timetable is an information entity. Information entities are generically dependent continuants. I disagree with the definition:  A set of facts...   it is not about actual facts, because the timetable is also a timetable if the bus company is on strike.'.
\item \emph{C--Expressing uncertainty about the decision of BFO class:} `Maybe processual context?'
\item \emph{D--General comment on BFO:} `Parts of the world that you would vacation to, such as Paris or a particular resort, are sites in BFO, but there's plenty of room for confusion with BFO spatial regions'.
\item \emph{E-- Possible alternative answer:} `Could also be site, as beach is ambiguous (1. material object; 2. a geographic region incliding 1.)'.
\item \emph{F--Comment on the survey itself:} `this is feeling like a lot more than 20 questions :-(The airspace tomorrow is an airspace. An airspace is a site. The temporal aspect is captured not in category but as time index in a relation predicate'.
\end{description}

From this, we note that, while there are some discussions about the survey (A and F), the majority of the comments were expressions of the difficulties of the task or justification for the decision made. Overall, we think that is supportive of our methodology and suggests that there are no systematic flaws in the task as specified, with the main factor being the underlying difficulty of the classification task.

\subsection{Assumptions and confounding factors}
While we believe our results are strong and provide an understanding of the reliability of \bfo experts to classify domain entities correctly underneath \bfo classes, there are also difficulties with the experiment and its design. We describe and discuss these below.

\begin{itemize}
\item We asked our participants to read a description of an entity and then place that entity within \bfo. Typically an ontologist would discuss an entity with a domain expert to understand that entity before deciding on the entity's classification. We used the travel domain as a source of entities as it is one that should be broadly familiar to all participants. Despite this mitigation, the nature of this task is a difficulty.
\item We wrote our descriptions in such a way as to try and make it clear as to the nature and identity of the entity, without writing the description in such a way that it told the participant the answer. This meant the descriptions were deliberately not definitional. An example is an entity such as `bus driver'; the wording could imply both an object and a role played by that object. The description has to be written in such a way to make it obvious which one is intended without saying directly which \bfo category is intended. This may have made some of the descriptions unclear or contrived, but our pilot studies were used to mitigate this factor.
\item The potential impact of survey fatigue was discussed in detail in Section~\ref{sec:time}.
\end{itemize}
We also have made some assumptions before setting up our study.

\begin{itemize}
\item As \bfo has a realist approach there is an assumption that any given entity will necessarily always go into the same place within \bfo.
\item The classification of any entity is independent of the motivation for developing the ontology. The motivation may determine the entities going into the ontology, but not where they are placed.
\item We've assumed that all BFO experts have the same commitment to BFO's worldview.
\end{itemize}

We had 8 participants from the list of \bfo~1.1 experts. Excluding the people we approached to pilot the study material, there was a pool of 16 people. So the number participating was half of the potential pool. Eight people may be considered to be low; nevertheless, we consider the results to be reasonably strong.

\section{Conclusion}

We find that classifying domain entities correctly under BFO, even for a well known general knowledge domain such as travel, is difficult even for BFO experts. On average, 8 experts had 2.65 different opinions about the correct classification of a domain entity--a result that could be an underestimate of the wider population, considering that the participants were all BFO experts and constitute a comparatively homogeneous sub-group of the wider group of ontology engineers. We do not see a reason to believe that the picture would be much different for other upper ontologies with comparable levels of abstraction.

We have found some indication that two branches of the BFO are of particular difficulty: \con{spatiotemporal region} and \con{spatial region}. It became apparent that these, in particular, need a clear methodological approach to ensure more consistent classifications and therefore less erroneous upper ontology integrations.

We confirmed that our participants were indeed aware when a classification was particularly difficult: the self-rated confidence after the fact correlated strongly with most of our correctness metrics.

We identified which classes were used often across a well-known repository of ontologies, the OBO Foundry, and showed that in general, there is no strong correlation between widespread use and classification correctness. One implication is, for example, that we should not expect the error rate for classifying domain entities under a BFO class to drastically decline once people start using that class more often.

We are aware, of course, that our methodology has some potential weaknesses. Participant numbers were relatively small, being bounded by the number of BFO experts in the world and those willing to take part. Some of our statistics are based on our own classification decision and we could just be wrong; we mitigated against this by prototyping our questionnaire (RS and PL defined the questions, and then trialed them on JM), as well as piloting against a number of BFO experts who were then not included in the main test. Furthermore, it was not ideal to present the questions in a fixed order to participants, as fatigue effects could have decreased classification performance towards the end of the survey. There is, indeed, evidence that suggests that more errors were made in the second half of the study than in the first: For example, the mean experimenter-expert similarity (EES) in the first half was 0.81, and in the second half 0.61. Finally, we have included statistics that are not dependent on our classifications. We believe that we have taken all reasonable steps to ensure the accuracy of our survey results.

We also note that the intention of this paper was not only to understand the difficulty of BFO integration \emph{per se} but to understand the difficulty of applying an upper (and potentially middle-level) ontology in general. We chose BFO because it is a widely used ontology, not because we think it is especially hard to use. We think that the same experiment done with other upper ontologies would also lead to the conclusion that they are hard to use and could benefit from some less craft-based methodological support. We think that our application of this form of a questionnaire is a methodology that could have broad use within the area of exploring methods to support the use of upper ontologies. If the use of upper ontologies is to become widespread in ontology engineering, we need methods to support that use; and to do that we need to understand the difficulties of using those upper ontologies.

Our study tells us little about how to improve the performance of BFO experts, let alone `regular' ontology authors, so what follows is speculation. Training in BFO, as well as other upper ontologies, should be a practical exercise. This means concentrating on the means by which an entity is properly classified, rather than on the philosophical underpinnings of a particular upper ontology's perspective. The key point is to convey an understanding of an upper ontology classes' characteristics, such that a recognition can be made as to where a particular domain entity should be placed. Correlation between correctness and usage was merely moderate in this study, but training could concentrate upon those parts of BFO that are most widely used, thus maximising the chance of correctly classifying under BFO, with an emphasis on those parts of the BFO that appear to cause the most difficulties. Finally, one of the utilities of an upper ontology is to guide ontologists as to which property to use; emphasising upper ontology use from this perspective should be beneficial.

Our results establish the need for the development of effective instruments and methods that support ontology engineers at integrating upper ontologies. One technique that we currently investigate to fill that need is the `twenty questions' approach~\cite{cooke1999knowledge,DBLP:journals/iwc/Welbank90} that comes from the area of knowledge elicitation. This technique can be used to establish the sequence of major characteristics of an entity that enable it to be recognised as belonging to a particular category; such a resource could be used to create a decision support instrument based upon the knowledge elicited from a population of upper ontology experts. We think that an empirically well-grounded set of questions tied to a particular upper ontology such as BFO could guide ontologists to the right class, significantly lowering the risk of highly inconsistent classifications. Any such technique needs to be at least as good as experts in using an upper ontology and be aware of issues in classifying entities within that upper ontology. We think that this study is a first step in establishing the empirical foundations for developing such a technique.

\textbf{Acknowledgments:}
We are grateful for the participation of the \bfo experts in this study. The contributions of NM and RS to this research have been funded by the EPSRC project: \emph{WhatIf: Answering ``What if\ldots'' questions for Ontology Authoring}. EPSRC reference EP/J014176/1.

\bibliographystyle{abbrvnat}
\bibliography{bfo_20q}

%
%
%
%
%

\end{document}